\documentclass[11pt,twoside]{article}
\usepackage{graphicx}
\usepackage{subfigure}
\usepackage{fancyheadings}
\usepackage{amsmath}
\usepackage{amssymb}
\usepackage{cite}
\usepackage{color,colortbl}

\definecolor{darkishgreen}{RGB}{39,203,22}
\definecolor{LightCyan}{rgb}{0.88,1,1}
\definecolor{Gray}{gray}{0.9}
\definecolor{lightRed}{RGB}{230,170,150}
\definecolor{modRed}{RGB}{230,82,90}
\definecolor{strongRed}{RGB}{230,6,6}

\begin{document}
\newcommand{\pst}{\hspace*{1.5em}}

\newcommand{\rigmark}{\em Journal of Russian Laser Research}
\newcommand{\lemark}{\em Volume 30, Number 5, 2009}

\newcommand{\be}{\begin{equation}}
\newcommand{\ee}{\end{equation}}
\newcommand{\bm}{\boldmath}
\newcommand{\ds}{\displaystyle}
\newcommand{\bea}{\begin{eqnarray}}
\newcommand{\eea}{\end{eqnarray}}
\newcommand{\ba}{\begin{array}}
\newcommand{\ea}{\end{array}}
\newcommand{\arcsinh}{\mathop{\rm arcsinh}\nolimits}
\newcommand{\arctanh}{\mathop{\rm arctanh}\nolimits}
\newcommand{\bc}{\begin{center}}
\newcommand{\ec}{\end{center}}

\thispagestyle{plain}

\label{sh}


\begin{center} {\Large \bf
\begin{tabular}{c}
DEFORMED ENTROPIC AND INFORMATION INEQUALITIES
\\[-1mm]
FOR $X$ - STATES OF TWO-QUBIT AND SINGLE QUDIT STATES
\end{tabular}
 } \end{center}

\bigskip

\bigskip

\begin{center} {\bf
V.I. Man'ko$^{1}$ and L.A. Markovich$^{2*}$
}\end{center}

\medskip

\begin{center}
{\it
$^1$P.N. Lebedev Physical Institute, Russian Academy of Sciences\\
Leninskii Prospect 53, Moscow 119991, Russia

\smallskip

$^2$Institute of Control Sciences, Russian Academy of Sciences\\
Profsoyuznaya 65, Moscow 117997, Russia
}
\smallskip

$^*$Corresponding author e-mail:~~~kimo1~@~mail.ru\\
\end{center}

\begin{abstract}\noindent
The $q$-deformed entropies of quantum and classical systems are discussed. Standard and $q$-deformed entropic inequalities
for $X$ - states of the two-qubit and a state of single qudit   with $j=3/2$ are presented.
\end{abstract}

\medskip

\noindent{\bf Keywords:}
$q$-deformed entropy, Entropic inequalities, $X$ - states, entanglement, noncomposite systems.
\section{Introduction}
\pst
The quantum correlations of bipartite qudit systems are characterized, e.g. by entropic inequalities written for von Neuman entropies \cite{Neuman} of the system and its subsystems \cite{Horodecki}.
The $q$-deformed entropies were introduced in \cite{Renyi,Tsallis}. These entropies being the functions of extra parameter contain more detailed information on properties
of density matrices of the qudit states and the qudit subsystem states. The Tsallis  entropy of bipartite qudit system was shown to satisfy the generalized subadditivity condition \cite{Audenaert,Petz}. This condition is the inequality available for Tsallis entropy of the bipartite system state and Tsallis entropies of two subsystem states. In approach \cite{Chernega,Chernega:14,Manko,OlgaMankoarxiv,Markovich3} it was shown that the relations for composite system state
can be extended to be valid for noncomposite systems, e.g. for the single qudit state. These inequalities reflect some quantum correlation  properties of degrees
of freedom either of subsystems (in the case of bipartite system) or degrees of freedom of the single qudit in the case of noncomposite system states.
One of the important states of the two-qubit systems are $X$-states. The properties of this states were studied for example in \cite{Markovich3,Mazhar,Hedemann}. The partial case of $X$-state is the Werner state \cite{Werner}.
The entanglement  properties of the Werner state were studied in detail for example in \cite{Lyons}.
\par The aim of our work is to obtain new deformed entropic inequality for $X$-state of composite (bipartite) and noncomposite (single qudit with $j=3/2$) quantum systems.
We consider both Renyi  and Tsallis  entropic inequalities.
\par The paper is constructed as follows. In Sec.~\ref{sec:1} we review the notion of Renyi and Tsallis entropies for bipartite systems. In Sec.~\ref{sec:2}  we obtain the new Tsallis entropic inequalities for $X$-state of noncomposite quantum system. The latter entropic inequality  is illustrated on the example of the Werner state of the single qudit.
\section{Renyi and Tsallis entropies}\label{sec:1}
\pst
Let us introduce the quantum state in the Hilbert space $\mathcal{H}$ defined by the following density matrix
\begin{eqnarray}\label{1}\rho&=&\left(
                                 \begin{array}{cccc}
                                   \rho_{11}& \rho_{12}& \rho_{13}& \rho_{14}\\
                                   \rho_{21}& \rho_{22}& \rho_{23}& \rho_{24}\\
                                   \rho_{31}& \rho_{32}& \rho_{33}& \rho_{34}\\
                                   \rho_{41}& \rho_{42}& \rho_{43}& \rho_{44}\\
                                 \end{array}
                               \right),\quad Tr(\rho)=1,\quad \rho=\rho^{\dagger},\quad\rho\geq0.
                               \end{eqnarray}
If we apply the invertible map of indices $1\leftrightarrow 1/2~1/2$; $2\leftrightarrow1/2~-1/2$; $3\leftrightarrow-1/2~1/2$; $4\leftrightarrow-1/2~-1/2$, the latter matrix can describe the two-qubit state. This provide the possibility to
construct reduced density operators $\rho_1=Tr_2\rho(1,2)$ and $\rho_2=Tr_1\rho(1,2)$ which describe the states of the subsystems $1$ and $2$, respectively. Applying another invertible map of indices $1\leftrightarrow 3/2$, $2\leftrightarrow1/2$, $3\leftrightarrow-1/2$, $4\leftrightarrow-3/2$, the density matrix \eqref{1} can be rewritten so, that it can describe the noncomposite system of the single qudit with $j=3/2$. Hence it is possible to use the density matrix in the form \eqref{1} to describe both bipartite systems as well as systems without subsystems. This idea to use invertible map of integers $1,2,3\ldots$ onto the pairs (triples, etc) of integers $(i,k)$, $j,k=1,2,\ldots$ to formulate the quantum properties of systems without subsystems was applied in \cite{Chernega,Chernega:14,Manko,OlgaMankoarxiv,Markovich3}. That gives us possibility to translate known properties of quantum correlations associated with structure of bipartite system like entanglement to the system without subsystems, e.g. single qudit.
\par An important measure of entanglement is entropy. The most known is the von Neumann entropy. It is obtained as
\begin{eqnarray*}S_N &=&-Tr\rho \ln\rho.\end{eqnarray*}
 More flexible are  Tsallis  and Renyi  entropies. The Renyi entropy generalizes the Shannon entropy, the Hartley entropy, the min-entropy, and the collision entropy. Both, Tsallis  and Renyi entropies, depend on extra parameter $q$, thus they  are called $q$-entropies. The classical $q$-entropies for the probability vector, constructed from the diagonal elements of the density matrix \eqref{1}  $\overrightarrow{p}=(p_1=\rho_{11},p_2=\rho_{22},p_3=\rho_{33},p_4=\rho_{44})$, are
 \begin{eqnarray}\label{2}S_{q}^{T}&=&\frac{1}{1-q}\left(\sum\limits_{i=1}^{4}p_i^q-1\right),\quad
 S_q^{R} =\frac{1}{1-q}\ln\left(\sum\limits_{i=1}^{4}p_i^q\right).
\end{eqnarray}
When $q\rightarrow1$, $S_{q}^{T}$ reduces to the von Neumann entropy. Tsallis and Renyi entropies can be written in the following form
 \begin{eqnarray*}S_{q}^{T}&=&-Tr\rho\ln_q\rho, \quad S_{q}^R =\frac{1}{1-q}\ln\left(Tr\rho^q\right)
 \end{eqnarray*}
 where
 \begin{eqnarray}\label{4}\ln_q\rho&=& \left\{
\begin{array}{ll}
\frac{\rho^{q-1}-I}{q-1}, &   \mbox{if}\qquad q\neq1,
\\
\ln\rho, & \mbox{if}\qquad q=1,
\end{array}
\right.
 \end{eqnarray}
for any real $q>0$, $I$ is identity matrix.  The logarithm \eqref{4} is called $q$-logarithm or deformed logarithm. The relations between Tsallis and Renyi entropies are given by the following formulas
 \begin{eqnarray}\label{11}S_{q}^{T}&=&\frac{\exp(S_{q}^R(1-q))-1}{1-q},\quad S_{q}^R=\frac{\ln(1+(1-q)S^T_{q})}{1-q}.
 \end{eqnarray}
\par If the density matrix \eqref{1} describes the  bipartite state (two-qubit system), then we can consider two subsystems on spaces $\mathcal{H}^{1}$ and $\mathcal{H}^{2}$ such that
$\mathcal{H}=\mathcal{H}^1\otimes\mathcal{H}^2$. Reduced density matrices $\rho_1$, $\rho_2$ are defined as partial traces  of \eqref{1}. Result matrices are density matrices of the density operators acting on  spaces $\mathcal{H}^1$ and $\mathcal{H}^2$, respectively.
Thus the reduced density matrices of the first and the second qubit are defined as
\begin{eqnarray}\rho_1={\left(
                              \begin{array}{cc}
                                \rho_{11}+\rho_{22} & \rho_{13}+\rho_{24} \\
                               \rho_{31}+\rho_{42} &  \rho_{33}+\rho_{44} \\
                              \end{array}
                            \right),\quad
                             \rho_2=\left(
                              \begin{array}{cc}
                                \rho_{11}+\rho_{33} & \rho_{12}+\rho_{34} \\
                               \rho_{21}+\rho_{43} &  \rho_{22}+\rho_{44} \\
                              \end{array}
                            \right)}\,. \label{3}
\end{eqnarray}
It is well known, that the von Neumann entropy is subadditive.
In \cite{Audenaert} was proved the subadditivity of the Tsallis entropy for $q>1$ for the composite system
 \begin{eqnarray*}S_{q}^{T}(\rho)&\leq&S_{q}^{T}(\rho_1)+S_{q}^{T}(\rho_2).
\end{eqnarray*}
There are other entropic inequalities, for example strong subadditivity condition \cite{Lieb}, which holds for the von Neumann entropy of three-partite quantum system. The fact that Tsallis entropy is not strong subadditive was recently proved in \cite{Petz}.
Let us define the $q$-information as
 \begin{eqnarray}\label{5}I_{q}^{T}&=&S_{q}^{T}(\rho_1)+S_{q}^{T}(\rho_2)-S_{q}^{T}(\rho)\geq0.
\end{eqnarray}
The subadditivity condition for the Tsallis entropy provides the inequality for the Renyi entropy
 \begin{eqnarray}\label{7}\exp(S_{q}^{R}(\rho_1)(1-q))+\exp(S_{q}^{R}(\rho_2)(1-q))-\exp(S_{q}^{R}(\rho)(1-q))<1.
\end{eqnarray}
Since Renyi and Tsallis entropies  tend to the von Neumann entropy for $q\rightarrow1$, both inequalities \eqref{5} and \eqref{7} in this limit give the standard positivity condition of the von Neumann mutual information.
\section{The Tsallis entropy for the $X$-state}\label{sec:2}
\begin{figure}[ht]
\begin{center}
\begin{minipage}[ht]{0.90\linewidth}
\includegraphics[width=1\linewidth]{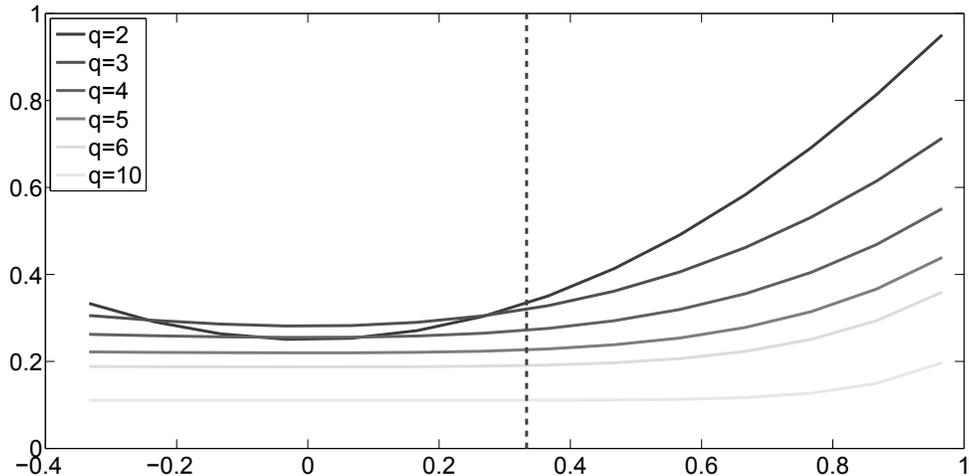}
\vspace{-4mm}
\caption{The $q$-information $I_{q}^{T}$ for the Werner state  \eqref{10} of the qudit with spin $j=3/2$ against the parameter $p$ for different values of the deformation parameter $q$.}
\label{fig:1}
\end{minipage}
\end{center}
\end{figure}

\pst
Using the invertible mapping  $1\leftrightarrow 3/2$, $2\leftrightarrow1/2$, $3\leftrightarrow-1/2$, $4\leftrightarrow-3/2$, the density matrix \eqref{1} can be rewritten as
\begin{eqnarray*}\rho_{3/2}&=&\left(
                                 \begin{array}{cccc}
                                   \rho_{3/2,3/2}& \rho_{3/2,1/2}& \rho_{3/2,-1/2}& \rho_{3/2,-3/2}\\
                                   \rho_{1/2,3/2}& \rho_{1/2,1/2}& \rho_{1/2,-1/2}& \rho_{1/2,-3/2}\\
                                   \rho_{-1/2,3/2}& \rho_{-1/2,1/2}& \rho_{-1/2,-1/2}& \rho_{-1/2,-3/2}\\
                                   \rho_{-3/2,3/2}& \rho_{-3/2,1/2}& \rho_{-3/2,-1/2}& \rho_{-3/2,-3/2}\\
                                 \end{array}
                               \right).
                               \end{eqnarray*}
This matrix is a density matrix of the single qudit state with spin $j=3/2$. Such system has no subsystems, thus it is impossible to write the reduced density matrices for it.
But using the form \eqref{1} we can successfully write them.
If $\rho_{12}=\rho_{13}=\rho_{21}=\rho_{31}=\rho_{24}=\rho_{34}=\rho_{42}=\rho_{43}=0$ then the density matrix \eqref{1} has the view of $X$-state density matrix
                               \begin{eqnarray}\label{2}
\rho^{X}&=&\left(
                     \begin{array}{cccc}
                       \rho_{11} & 0 & 0 & \rho_{14}\\
                       0 & \rho_{22}& \rho_{23} & 0 \\
                       0 & \rho_{32} & \rho_{33} & 0 \\
                       \rho_{41} & 0 & 0 & \rho_{44} \\
                     \end{array}
                   \right)=\left(
                     \begin{array}{cccc}
                       \rho_{11} & 0 & 0 & \rho_{14}\\
                       0 & \rho_{22}& \rho_{23} & 0 \\
                       0 & \rho_{23}^{\ast} & \rho_{33} & 0 \\
                      \rho_{14}^{\ast} & 0 & 0 & \rho_{44} \\
                     \end{array}
                   \right),
\end{eqnarray}
where $\rho_{11},\rho_{22},\rho_{33},\rho_{44}$ are positive reals and $\rho_{23},\rho_{14}$ are complex quantities.
The latter matrix has the unit trace and is nonnegative if $\rho_{22}\rho_{33}\geq|\rho_{23}|^2$, $\rho_{11}\rho_{44}\geq|\rho_{14}|^2$.
The reduced density matrices are defined as
\begin{eqnarray*}\rho_1&=&\left(
                            \begin{array}{cc}
                              \rho_{11}+\rho_{22} & 0 \\
                              0 & \rho_{33}+\rho_{44} \\
                            \end{array}
                          \right),\quad
                          \rho_2=\left(
                            \begin{array}{cc}
                              \rho_{11}+\rho_{33}& 0 \\
                              0 & \rho_{22}+\rho_{44} \\
                            \end{array}
                          \right).
\end{eqnarray*}
Hence, the $q$-information \eqref{5} for the $X$-state of the single qudit is
\begin{eqnarray}\label{6}I_{q}^{T}&=&\frac{1}{1-q}\Bigg(
(\rho_{11} + \rho_{22})((\rho_{11} + \rho_{22})^{q - 1} - 1)
+ (\rho_{11} + \rho_{33})((\rho_{11} + \rho_{33})^{q - 1} - 1)\\\nonumber
&+& (\rho_{22} + \rho_{44})((\rho_{22} + \rho_{44})^{q - 1} - 1)+
 (\rho_{33} + \rho_{44})((\rho_{33} + \rho_{44})^{q - 1} - 1)\\\nonumber
 &-& (\rho_{11}+\rho_{22}+\rho_{33}+\rho_{44})(\rho^{q - 1} - 1)\Bigg)\geq0.
 \end{eqnarray}
 As an example of the $X$-state density matrix of the qudit state with spin $j=3/2$ can be taken the Werner state matrix
\begin{eqnarray}\label{10}\rho^{W}&=&\left(
                     \begin{array}{cccc}
                       \frac{1+p}{4} & 0 & 0 & \frac{p}{2}\\
                       0 & \frac{1-p}{4}& 0 & 0 \\
                       0 & 0& \frac{1-p}{4} & 0 \\
                       \frac{p}{2} & 0 & 0 & \frac{1+p}{4} \\
                     \end{array}
                   \right),
\end{eqnarray}
where parameter is $-\frac{1}{3}\leq p\leq1$. The parameter domain $\frac{1}{3}< p\leq1$ corresponds to the entangled state.
The information \eqref{6} for the latter state can be seen in Fig.~\ref{fig:1}.
The dashed line in the point $p=\frac{1}{3}$ marks the border between the separable  and entangled Werner states. One can see general behavior of the $q$-information against parameter $p$ for different values of deformation parameter $q$. In the domain of the entangled states the $q$-information increases with increasing the degree of the state entanglement.
The sensitivity of the $q$-information to the degree of entanglement depends on the deformation parameter $q$.
\section{Summary}
\pst
To conclude we point out the main results of the work. We applied $q$-deformed entropies of Renyi and Tsalles as a measure of entanglement for the systems without subsystems.
New deformed entropic inequality for the $X$-state of noncomposite quantum state (qudit with spin $j=3/2$) was written. As an example of the $X$-state the Werner state with one parameter was taken.
\par Despite there are no subsystems in such systems it is possible
to introduce analogs of partial traces like for composite systems using special mapping illustrated in the text. Certainly it is necessary to understand that these partial traces for noncomposite systems have not the same meaning as for the composite system quantum state. The understanding of physical background of the correlations inside the system without subsystems is still not well understood and  will be developed in future article of the authors.

\end{document}